\documentclass[12pt]{article}
\newcommand{\be}{\begin{equation}}
\newcommand{\ee}{\end{equation}}
\newcommand{\beqa}{\begin{eqnarray}}
\newcommand{\eeqa}{\end{eqnarray}}

\begin{document}
\vskip1cm
\begin{flushright} 
\end{flushright}
\vskip1cm
\begin{center}
{\Large\bf New concept of relativistic invariance\\ in NC
space-time: twisted Poincar\' e symmetry and its implications}
\vskip1cm {\bf M. Chaichian, P. Pre\v{s}najder\footnote{On leave
of absence from the Department of Theoretical Physics and Physics
Teaching, Comenius University, Mlynsk\'{a}
dolina, SK-84248 Bratislava, Slovakia} and A. Tureanu}\\
High Energy Physics Division, Department of Physical Sciences,\\
University of Helsinki, and Helsinki Institute of Physics\\ P.O.
Box 64, FIN-00014
Helsinki, Finland\\
\end{center}
\vskip1cm

{\bf Abstract}: We present a systematic framework for
noncommutative (NC) QFT within the new concept of relativistic
invariance based on the notion of twisted Poincar\' e symmetry
(with all 10 generators), as proposed in ref. \cite{CKT}. This
allows to formulate and investigate all fundamental issues of
relativistic QFT and offers a firm frame for the classification of
particles according to the representation theory of the twisted
Poincar\' e symmetry and as a result for the NC versions of CPT
and spin-statistics theorems, among others, discussed earlier in
the literature. As a further application of this new concept of
relativism we prove the NC analog of Haag's theorem.
\vskip1cm

\section{Introduction}

The idea that coordinates may not commute can be traced back to
Heisenberg. For an early reference on field theory in a
noncommutative space-time, see \cite{snyder}. The noncommutative
(NC) QFT in Moyal space-time was formulated in \cite{doplicher}.
Within this approach the fields are functions of NC coordinates
${\hat  x}^\mu$, $\mu=0,1,2,3$, satisfying the commutation
relations
\be [{\hat x}^\mu,{\hat x}^\nu]\ =\ {\rm i}\theta^{\mu\nu}\ ,\
\theta^{\mu\nu}=-\theta^{\nu\mu}\ -\ {\rm real\ constants}\
.\label{I.1}\ee
This type of theory emerges also as a low energy limit of the
string theory in a particular background, \cite{seibergwitten}. It
is well-known that if there is space-time noncommutativity, i.e.
some of $\theta^{0i}\neq 0$, $i=1,2,3$, the resulting theory
violates causality and unitarity \cite{SST}, \cite{GM}. Therefore,
we shall consider only space-space noncommutativity with a choice
of space coordinates such that $\theta^{12}=-\theta^{21}\equiv
\theta$ with all other $\theta^{ij}$'s vanishing. The commutation
relations (\ref{I.1}) preserve the translational symmetry but
break the Lorentz invariance. With the choice $\theta^{12}\neq 0$
the Lorentz symmetry is broken down to the {\it residual}
$SO(1,1)\times SO(2)$ {\it symmetry} formed by boost in 3-rd space
direction and rotations in (1,2)-plane.

An axiomatic formulation of NC QFT, based on this residual
space-time symmetry, was proposed in \cite{AG}. The resulting
theory is effectively (1+1)-dimensional QFT in (0,3)-Minkowski
plane supplemented by instantaneous interactions in (1,2)
Euclidean plane. The methods of usual (commutative) relativistic
QFT are applied in (0,3)-Minkowski plane, whereas the role of NC
coordinates ${\hat x}^1$ and ${\hat x}^2$ is strongly suppressed.
This enables to investigate {\it some} general aspects which can
be adapted to the residual space-time symmetry. However, this
approach cannot be considered to be satisfactory: (i) There are no
obvious reasons why the full relativistic symmetry should emerge
in the commutative limit $\theta\to 0$, and (ii) there is no
natural way how to introduce the spin and the classification of
particles.

Recently, in \cite{CKT} was proposed an alternative interpretation
of commutation relations (\ref{I.1}). Using the notion of twisted
Poincar\' e symmetry (with {\it all} ten generators $P_\mu$ and
$M_{\mu\nu}$) the interpretation of (\ref{I.1}) was extended from
Lie algebra framework to Hopf algebras: {\it If in the usual
(commutative) case, relativistic invariance means symmetry under
the Poincar\' e transformations, in the noncommutative case
relativistic invariance means symmetry under twisted Poincar\' e
transformations}. This "hidden" symmetry enables us to discuss all
the aspects of relativistic QFT, which are not accessible within
the residual symmetry approach. For example, it justifies the
attempt to prove the spin-statistics theorem in \cite{CNT} in
Lagrangian formulation and in \cite{CMNTV} within the axiomatic
approach.

The main aim of this paper is to present a consistent frame for NC
QFT with the relativistic invariance realized in terms of the
twisted Poincar\' e symmetry proposed previously and to apply the
developed techniques to the investigation and proof of the Haag
theorem \cite{Haag}, with the statement: If a field at a certain
time is related to a free one by a unitary transformation, as is
the case in the interaction picture, then the field is inevitably
free. This theorem concerns deep mathematical subtleties of
systems possessing an infinite number of degrees of freedom, as is
the case of QFT. It is an ultimate question to understand its
background and nature also in the framework of NC QFT. Since the
latter theories are nonlocal, one might  think that the analog of
Hagg's theorem may no more be valid due to the nature of NC field
theories: while interacting fields are nonlocal, the free fields
are local.

In Section 2 we shall first present the basic notions defining the
twisted Poincar\' e symmetry, and then we perform the twist
deformation of the QFT operator algebra generated by powers of
quantum fields. Section 3 is devoted to the proof of Haag's
theorem in the twisted Poincar\' e symmetry approach. Section 4
contains the concluding remarks.

\section{Twisted NC QFT}

{\bf Twisted Poincar\'{e} symmetry}. First we briefly describe the
twisted Poincar\'{e} Hopf-algebra ${\cal U}_t({\cal P})$ proposed
recently in \cite{CKT} (for basic notions see \cite{CD}):

(a) The usual Poincar\'{e} algebra ${\cal P}$ is spanned by the
generators of translations $P_\mu$ and Lorentz transformations
$M_{\mu\nu}= -M_{\nu\mu}$, $\mu,\nu=0,1,2,3$, satisfying the usual
commutation relations. Its enveloping algebra ${\cal U}({\cal P})$
is a so called trivial Hopf-algebra with the co-product
$\Delta(X)=X\otimes 1+1\otimes X$.

(b) An essential object for us is the twisted Poincar\'{e}
Hopf-algebra ${\cal U}_t({\cal P})$. In ${\cal U}_t({\cal P})$ the
co-product is defined by $\Delta_t(X)={\cal F}\Delta(X){\cal
F}^{-1}$ 
with the twist given as
\be {\cal F}=e^{\frac{{\rm i}}{2}\theta^{\mu\nu}P_\mu\otimes
P_\nu}\ ,\ P_\mu(X)\ \equiv\ [P_\mu,X]\ .\label{I.2}\ee
Using ${\theta_\mu}^\rho= \eta_{\mu\sigma}\theta^{\sigma\rho}$,
equation (\ref{I.2}) implies (see \cite{CKT}, \cite{wess}):
\[ \Delta_t(P_\mu)\ =\ P_\mu\otimes 1\ +\ 1\otimes P_\mu\ ,\]
\[ \Delta_t(M_{\mu\nu})\ =\ M_{\mu\nu}\otimes 1\ +\ 1\otimes
M_{\mu\nu} \]
\be -\ \frac{1}{2}({\theta_\mu}^\rho P_\nu-{\theta_\nu}^\rho
P_\mu)\otimes P_\rho\ +\
\frac{1}{2}P_\rho\otimes({\theta_\mu}^\rho P_\nu-{\theta_\nu}^\rho
P_\mu)\ .\label{I.3}\ee
The antipode (the reverse) and co-unit are defined in ${\cal
U}({\cal P})$ and ${\cal U}_t({\cal P})$ in the standard way; we
shall not need them explicitly.

Having the representation of a Hopf-algebra in an associative
algebra $\hat{\cal A}$ compatible with the co-product (Leibniz
rule), the multiplication $m({\hat a}\otimes{\hat b})\equiv{\hat
a}{\hat b}$ in $\hat{\cal A}$ has to be changed after a twisting
according to (\ref{I.2}). In our particular case the twisted
multiplication ($\star$-product) reads \cite{CKT}:
\be {\hat a}\star{\hat b}\ \equiv\ m_t\left({\hat a}\otimes{\hat
b}\right)\ =\ m\left(e^{-\frac{{\rm i}}{2} \theta^{\mu\nu}{\hat
P}_\mu\otimes{\hat P}_\nu}{\hat a}\otimes{\hat b}\right)\ ,\ {\hat
P}_\mu ({\hat a})\equiv[{\hat P}_\mu,{\hat a}]\ ,\label{I.4}\ee
where ${\hat P}_\mu$ is the representation of $P_\mu$ in
$\hat{\cal A}$, and similarly ${\hat M}_{\mu\nu}$ denotes the
representation of $M_{\mu\nu}$.

{\bf Algebra of quantum fields}. Below we shall apply this
abstract concept to the algebra $\hat{\cal A}$ of operators in QFT
Hilbert space ${\cal H}$ generated by products of fields
${\hat\phi} (x_1)\dots{\hat\phi}(x_n)$, where $x_1,\dots,x_n$, are
space-time points in the Minkowski space $M$.

We shall assume that ${\cal H}$ carries a unitary representation
of Poincar\' e algebra ${\cal P}$ with self-adjoint generators
${\hat P}_\mu$ and ${\hat M}_{\mu\nu}$. As usual we assume that
${\hat P}_\mu$ has the spectrum  $V_+=\{p^\mu;\, p^0\geq|\vec{p}
|\}$. Furthermore, we assume that there is a unique ${\cal
P}$-invariant vacuum state $|0\rangle\in{\cal H}$: ${\hat
P}_\mu|0\rangle={\hat M}_{\mu\nu}|0\rangle =0$, and that
$|0\rangle$ is a cyclic vector in Hilbert space, i.e. ${\cal H}$
is spanned by vectors ${\hat\phi}(x_1)\dots{\hat\phi}(x_n)
|0\rangle$, $x_1,\dots,x_n\in M$.

The operators ${\hat P}_\mu$ and ${\hat M}_{\mu\nu}$ act on fields
in a standard way:
\[ [{\hat P}_\mu,{\hat\phi}(x)]\ =\ -{\rm i}\partial_\mu{\hat\phi(x)}\
\equiv\ ({\cal P}_\mu{\hat\phi})(x)\ ,\]
\be [{\hat M}_{\mu\nu},{\hat\phi}(x)]\ =\ {\rm i}(x_\mu
\partial_\nu-x_\nu\partial_\mu){\hat\phi}(x)\ \equiv\
({\cal M}_{\mu\nu}{\hat\phi})(x)\ .\label{I.6A}\ee
We extend this action via commutators to any product of fields.
For example, the Lorentz generator ${\hat M}_\omega=\frac{1}{2}
\omega^{\mu\nu}{\hat M}_{\mu\nu}$, $\omega^{\mu\nu}=
-\omega^{\nu\mu}$, acts on the product
${\hat\phi}(x){\hat\phi}(y)$ as follows:
\[ {\cal M}_{\mu\nu}({\hat\phi}(x){\hat\phi}(y))\ \equiv\
[{\hat M}_{\mu\nu},{\hat\phi}(x){\hat\phi}(y)]\ =\ [{\hat
M}_{\mu\nu},{\hat\phi}(x)]{\hat\phi}(y)+{\hat\phi}(x)[{\hat
M}_{\mu\nu},{\hat\phi}(y)] \]
\be =\ ({\cal M}_{\mu\nu}{\hat\phi})(x){\hat\phi}(y)\ +\
{\hat\phi}(x)({\cal M}_{\mu\nu}{\hat\phi})(y)\ .\label{I.6B}\ee
Equipped with this action, the algebra $\hat{\cal A}$ carries a
representation of the trivial Hopf algebra ${\cal U}({\cal P})$.

Deforming now the algebra $\hat{\cal A}$ by the twist (\ref{I.2}),
we arrive at the field operator algebra $\hat{\cal A}_\star$
carrying the representation of ${\cal U}_t({\cal P})$ with the
twisted operator product specified by (\ref{I.4}) and
(\ref{I.6A}):
\be {\hat\phi}(x)\star{\hat\phi}(y)\ =\ m\left(e^{-\frac{{\rm
i}}{2} \theta^{\mu\nu}{\hat P}_\mu\otimes{\hat
P}_\nu}{\hat\phi}(x) \otimes{\hat\phi}(y)\right)\ =\ e^{\frac{{\rm
i}}{2}\theta^{\mu\nu}
\partial_{x^\mu}\partial_{y^\nu}}{\hat\phi}(x){\hat\phi}(y)\
.\label{I.10}\ee
We would like to stress that although (\ref{I.10}) looks like a
Moyal product, there has been {\it no} noncommutativity of
coordinates used. The $\star$-product is {\it inherited} from the
twist of the operator product of quantum fields.

{\bf Locality condition}. The locality condition is an independent
axiom. It is the only place where the twist explicitly influences
the field axioms, since at this point we are dealing with the
products of fields.

Let us now investigate the transformation properties of
star-product of fields ${\hat\phi}(x)\star{\hat\phi}(y)$ under the
Lorentz transformation ${\hat M}_\omega=\frac{1}{2}\omega^{\mu\nu}
{\hat M}_{\mu\nu}$, $\omega^{\mu\nu}=-\omega^{\nu\mu}$. From
(\ref{I.3})-(\ref{I.6A}) it follows that the corresponding action
of the Lorentz generator is given by:
\[ {\cal M}^t_{\omega}({\hat\phi}(x)\star{\hat\phi}(y))\ \equiv\
m_t\left(\Delta_t({\hat M}_\omega){\hat\phi}(x) \otimes
{\hat\phi}(y)\right) \]
\[ =\ ({\cal M}_\omega{\hat\phi})(x)\star{\hat\phi}(y)\ +\
{\hat\phi}(x)\star({\cal M}_\omega{\hat\phi})(y) \]
\be -\ \frac{1}{2}\theta^{\rho\sigma}{\omega_\rho}^\nu({\cal
P}_\nu{\hat\phi})(x)\star({\cal P}_\sigma{\hat\phi})(y)\ -\
\frac{1}{2}\theta^{\rho\sigma}{\omega_\sigma}^\nu({\cal
P}_\rho{\hat\phi})(x)\star({\cal P}_\nu{\hat\phi})(y)\
.\label{I.6C}\ee
In spite of the $\theta$-corrections in the last line, the
operators ${\cal M}^t_{\omega}$ satisfy the Lorentz algebra
commutation relations:
\[ [{\cal M}^t_{\omega},{\cal M}^t_{\omega'}]({\hat\phi}(x)\star
{\hat\phi}(y))\ =\ {\rm i}{\cal M}^t_{\omega\times\omega'}
({\hat\phi}(x)\star{\hat\phi}(y))\ ,\]
where $(\omega\times\omega')^{\mu\nu}=\omega^{\mu\sigma}
{{\omega'}_\sigma}^\nu -\omega^{\prime\mu\sigma}
{\omega_\sigma}^\nu$. This follows from the equation
\be {\cal M}^t_{\omega}\ =\  e^{\frac{\rm i}{2}
\theta^{\mu\nu}\partial_{x^\mu}\partial_{y^\nu}}{\cal
M}^t_{\omega} e^{-\frac{\rm i}{2}
\theta^{\mu\nu}\partial_{x^\mu}\partial_{y^\nu}},\label{I.7}\ee
which simply reflects the fact that the twist in question
influences the co-product but {\it not} the basic algebraic
structure itself.

Thus, we can exponentiate the action of Lorentz generators in a
standard way, and obviously this is valid for translations too. In
the Appendix it is shown that this guarantees the {\it twisted
Poincar\'{e} covariance} of the $\star$-product of fields:
\be {\hat\phi}(x)\star{\hat\phi}(y)\ =\ e^{\frac{\rm i}{2}
\theta^{\mu\nu}\partial_{x^\mu}\partial_{y^\nu}} {\hat\phi}(x)
{\hat\phi}(y)\ \mapsto\ e^{\frac{\rm i}{2}\theta^{\mu\nu}
\partial_{x^{\prime\mu}}\partial_{y{\prime^\nu}}}{\hat\phi}(x')
{\hat\phi}(y')\ =\ {\hat\phi}(x')\star{\hat\phi}(y')\
,\label{I.8}\ee
where $x'=\Lambda x+a$ and $y'=\Lambda y+a$ are the accordingly
transformed space-time points. We stress that $\theta^{\mu\nu}$ is
a {\it twisted Poincar\' e invariant tensor}:
$\theta^{\mu\nu}=\theta^{\prime\mu\nu}$. This is natural and in
full agreement with \cite{CKT}.

The covariance property (\ref{I.8}) {\it dictates} the
relativistic form of locality condition among $\star$-products of
fields. We postulate it in the simplest form:
\be (x-y)^2=(x^0-y^0)^2-(\vec{x}-\vec{y})^2<0\ \Rightarrow\
[{\hat\phi}(x),{\hat\phi}(y)]_\star\ =\ 0\ .\label{I.11A}\ee

{\it Note}: In \cite{CT} was discussed, within NC context, a
weaker locality condition in the form
\be (x-y)^2\ <\ -l^2\ \Rightarrow\ [{\hat\phi}(x),
{\hat\phi}(y)]_\star\ =\ 0\label{I.11C}\ee
with $l^2\approx|\theta|$. In \cite{vladimirov} it was proven that
in a field theory satisfying {\it translational invariance} and
{\it spectral condition} $p\in V_+$, the locality condition
(\ref{I.11C}) implies (\ref{I.11A}). In our case, these
requirements are indeed satisfied, thus supporting the use of the
form (\ref{I.11A}) for the locality axiom.

\section{Haag's theorem}

{\bf Wightman functions}. The NC Wightman functions are defined as
the vacuum expectation values of multiple $\star$-products of
fields \cite{CMNTV}:
\[ W_\star(x_1,\dots,x_n)\ \equiv\
\langle0|{\hat\phi}(x_1)\star\dots\star{\hat\phi}(x_n)|0\rangle \]
\be =\ e^{\frac{\rm i}{2}\theta^{\mu\nu}\sum_{a<b}
\partial_{x^\mu_a}\partial_{x^\nu_b}}W(x_1,\dots,x_n)\ ,\label{III.1a}\ee
where $W(x_1,\dots,x_n)=\langle0|{\hat\phi}(x_1)\dots{\hat\phi}
(x_n)|0\rangle$. Due to translational invariance, the $\theta
$-dependent exponent in (\ref{III.1a}) can be omitted for the
2-point NC Wightman function: $W_\star(x,y)=W(x,y)={\cal W}(x-y)$.
However, it cannot be omitted for the higher NC Wightman
functions.

The NC Wightman functions defined above satisfy the twisted
Poincar\' e covariance condition and the twisted Poincar\' e
locality condition:
\[ W_\star(x_1,\dots,x_n)\ \mapsto\ W_\star(\Lambda
x_1+a,\dots,\Lambda x_n+a)\ ,\]
\be (x-y)^2\ <\ 0\ \Rightarrow\ W_\star(\dots,x,y,\dots)=
W_\star(\dots,y,x,\dots)\ ,\label{III.2a}\ee
together with the usual positivity axiom (the positivity of the
norm in ${\cal H}$).

This enables to formulate and discuss various fundamental NC QFT
theorems along the lines similar to the standard ones. The
approach based on twisted Poincar\' e symmetry can offer a firm
foundation for the proofs of NC analogs of the CPT theorem and
spin-statistics theorem in the axiomatic approach.

{\bf Haag's theorem}. As an interesting application of the twisted
Poincar\' e formalism we prove the Haag theorem. The theorem was
originally formulated in \cite{Haag}; its variations and
extensions were presented in \cite{FJ} (for a detailed discussion,
see \cite{pct}). Below we briefly discuss a sequence of theorems
leading, in the twisted Poincar\' e context, to the Haag theorem.
The first theorem we need can be formulated as follows:

{\bf Theorem 1}: {\it Let the 2-point function $\langle0|
{\hat\phi}(x)\star{\hat\phi} (y)|0\rangle$ of an arbitrary scalar
field ${\hat\phi}(x)$ coincide with the 2-point function of a free
field of mass $m>0$, $-{\rm i}\Delta(x-y;m)$, satisfying the
Klein-Gordon equation, so that
\be (\partial^2_x+m^2)\langle0|{\hat\phi(x)}\star{\hat\phi}
(y)|0\rangle\ =\ -{\rm i}(\partial^2_x+m^2)\Delta(x-y;m)\ =\ 0\
.\label{III.1}\ee
Then the interaction current vanish:} ${\hat J}(x)\equiv
(\partial^2_x+m^2)\hat\phi(x)=0$, {\it i.e. the theory is free}.

{\it Proof}: The first part of the proof is identical to the usual
one \cite{FJ}. Considering an arbitrary test function $F(x)$ we
shall prove that the state $|\psi\rangle\equiv\int dx\,F(x) {\hat
J}(x)|0 \rangle=\int dx\,F(x)(\partial^2_x+m^2)\hat\phi(x)
|0\rangle$ vanishes. Omitting the $\star$-product in (\ref{III.1})
we obtain
\[ \langle\psi|\psi\rangle\ =\ \int dxdy\,{\bar F}(x)F(y)
(\partial^2_x+m^2)(\partial^2_y+m^2)\langle0|\hat\phi(x)
\hat\phi(y)|0\rangle\ =\ 0\ .\]
Since $F(x)$ is arbitrary, the current ${\hat J}(x)$ annihilates
the vacuum state: ${\hat J}(x)|0\rangle$ $=0$.

The nontrivial part of the theorem is based on the existence of NC
analogs of Jost-points for the Wightman functions $W_\star
(x_1,\dots,x_{n-1},x_n)=\langle0|{\hat\phi}(x_1) \star\dots
\star{\hat\phi}(x_{n-1})\star{\hat\phi}(x_n)|0\rangle$ which are
boundary values of a holomorphic function $W_\star(z_1,\dots,
z_{n-1},z_n)$ of complex variables $z_1,\dots,z_{n-1},z_n$ in a
proper domain (the extended tube) containing real Jost points $(r_1,\dots,r_{n-1},r_n)$, satisfying
$(r_k-r_l)^2<0$, for all $k>l$. Due to the locality condition
(\ref{III.2a}), we can permute the fields evaluated at Jost
points:
$W_\star(r_1,\dots,r_{n-1},r_n)=W_\star(r_1,\dots,r_n,\dots,r_{n-1})$.
Performing now in this equation the analytical continuation back
to $x_1\dots,x_{n-1},x_n$, and acting by $(\partial^2_{x_n}+m^2)$
we obtain:
\be \langle0|\hat\phi(x_1)\star\dots\star\hat\phi(x_{n-1})\star
{\hat J}(x_n)|0\rangle\ =\ \langle0|\hat\phi(x_1)\star\dots\star
{\hat J}(x_n)\star\dots\star \hat\phi(x_{n-1})|0\rangle\
.\label{III.4}\ee
Since ${\hat J}(x)|0\rangle=0$, the l.h.s. of (\ref{III.4}) is
zero. The r.h.s. represents, due to the completeness axiom, an
arbitrary matrix element of the interaction current. Consequently,
${\hat J}(x)=0$, i.e. the theory is {\it free}.

To prove the Haag theorem we need a simple theorem dealing with
fixed time fields. Its formulation and proof is similar to the
usual one (see \cite{pct}).

{\bf Theorem 2}: {\it Let $\hat\phi_1(\vec{x},t)$ and
$\hat\phi_2(\vec{x},t)$ be two irreducible scalar fields at a
fixed time $t$ defined respectively in Hilbert spaces ${\cal H}_1$
and ${\cal H}_2$ in which there are two continuous unitary
representations of the Euclidean group $E(3)$:
\[ U_j(\vec{a},R)\hat\phi_j(\vec{x},t)U^\dagger_j(\vec{a},R)\ =\
\hat\phi_j(R\vec{x}+\vec{a},t)\ ,\ j=1,2\ .\]
We assume that: (i) The representations possess unique invariant
vacuum states $|0\rangle_j$: $U_j(\vec{a},R)$ $|0\rangle_j=
|0\rangle_j$, $j=1,2$, and (ii) there exists a unitary operator
$V:{\cal H}_1\to{\cal H}_2$ such that at time t:
$\hat\phi_2(\vec{x},t)=V\hat\phi_1(\vec{x},t)V^\dagger$.

Then} $U_2(\vec{a},R)=VU_1(\vec{a},R)V^\dagger$, $|0\rangle_2\ =\
e^{{\rm i}\alpha}V|0\rangle_1$, $\alpha\in{\bf R}^1$.

{\it Note}: The group $E(3)$ is a subgroup of the Poincar\' e
group generated by rotations and translations in ${\bf R}^3$. This
induces the representation of the twisted Euclidean Hopf-algebra
by twist (\ref{I.2}) in the operator algebra generated by field
operators at the fixed time $t$. It is a Hopf subalgebra of
$\hat{\cal A}_\star$ - this is the place where the conditions
$\theta^{0i}=0$, $i=1,2,3$, become essential.

{\bf Corollary}: The equal time vacuum expectation values of both
theories coincide: ${}_1\langle0|\hat\phi_1(\vec{x}_1,t)\star\dots
\star\hat\phi_1(\vec{x}_n,t)|0\rangle_1= {}_2\langle0|\hat\phi_2
(\vec{x}_1,t)\star\dots\star\hat\phi_2(\vec{x}_n,t)|0\rangle_2$.

Theorem 1 and the Corollary provide us with the proof of Haag's
theorem:

{\bf Theorem 3} (Haag): {\it Suppose $\hat\phi_1(x)$ is a free
hermitian scalar field of mass $m>0$, and $\hat\phi_2(x)$ is a
scalar field satisfying the NC QFT axioms given above. Suppose
further that the fields $\hat\phi_j(x)$,
$(\partial_t\hat\phi)_j(x)$, $j=1,2$, satisfy the hypothesis of
Theorem 2. Then $\hat\phi_2(x)$ is a free field of mass $m$}.

{\it Proof}: Any two space-like separated points $x$ and $y$ can
be brought by a Lorentz transformation to equal-time plane: $x=
(\vec{x},t)$ and $y=(\vec{y},t)$. By Corollary we have ${}_2
\langle0|\hat\phi_2(\vec{x},t)\star\hat\phi_2(\vec{y},t)|0\rangle_2=
-{\rm i}\Delta(\vec{x}-\vec{y},0;m)$. Using the standard analytic
continuation argument and the twisted relativistic covariance of
$\hat\phi_2(x)\star\hat\phi_2(y)$ we obtain: ${}_2\langle0|
\hat\phi_2(x)\star \hat\phi_2(y)|0\rangle_2= -{\rm i}\Delta(x-y;
m)$. The Haag's theorem is then a direct consequence of Theorem 1.

\section{Concluding remarks}

The approach presented above is based on the twisted Poincar\' e
symmetry (with all 10 generators) which states that: while
symmetry under the usual Lorentz transformations guarantees the
relativistic invariance of a theory, in the NC QFT the concept of
relativistic invariance, however, should be replaced by the
requirement of invariance of the theory under the twisted
Poincar\' e transformations. The latter allows us to formulate and
discuss all fundamental issues of relativistic QFT within NC
context. Because of the content of its representations, the
twisted Poincar\' e algebra offers a firm framework for the proofs
of the NC version of CPT and the spin-statistics theorems
\cite{CMNTV}, among other results obtained in the literature so
far, and justifies in particular the results obtained in \cite{CNT} within
Lagrangian formulation. The Haag theorem also belongs to that
class of exact (not perturbative) results.

\vskip 0.3cm {\bf{Acknowledgements}}. We are grateful to R. Haag,
P. Kulish and K. Nishijima for valuable discussions and remarks.
The financial support of the Academy of Finland under the Projects
No. 54023 and 104368 is greatly acknowledged. P.P.'s work was
partially supported by project VEGA No. 1/025/03. \vskip1cm

{\Large\bf Appendix}

Let us consider an infinitesimal Lorentz transformation
${\hat\phi}(x)\star{\hat\phi}(y)\mapsto(1+{\rm i}{\cal
M}^t_{\omega})({\hat\phi}(x)\star{\hat\phi}(y))$
generated by ${\hat M}_\omega=\frac{1}{2} {\omega^{\mu\nu}\hat
M}_{\mu\nu}$, $\omega$ - infinitesimal. According to the
definition (\ref{I.6C}) we have:
\[ (1+{\rm i}{\cal M}^t_{\omega})({\hat\phi}(x)\star
{\hat\phi}(y))\ =\ m_t\{[1+{\rm i}\Delta_t({\hat
M}_\omega)][{\hat\phi}(x)\otimes{\hat\phi}(y)]\} \]
\[ =\ m\{e^{-\frac{\rm i}{2}\theta^{\rho\sigma}{\cal P}_\rho\otimes
{\cal P}_\sigma}[{\hat\phi}(x)\otimes{\hat\phi}(y)+{\rm i}({\cal
M}_{\omega}{\hat\phi})(x)\otimes{\hat\phi}(y)+{\rm i}
{\hat\phi}(x)\otimes({\cal M}_{\omega}{\hat\phi}){\hat\phi}(y)\]
\[ -\ \frac{\rm i}{2}\theta^{\rho\sigma}{\omega_\rho}^\nu({\cal
P}_\nu{\hat\phi})(x)\otimes({\cal P}_\sigma{\hat\phi})(y)\ -\
\frac{\rm i}{2}\theta^{\rho\sigma}{\omega_\sigma}^\nu({\cal
P}_\rho{\hat\phi})(x)\otimes({\cal P}_\nu{\hat\phi})(y)]\}\ .\]
Up to terms linear in $\omega$ this can be rewritten as
\[ (1+{\rm i}{\cal M}^t_{\omega})({\hat\phi}(x)\star{\hat\phi}(y))
\ \doteq\ m\{e^{-\frac{\rm i}{2}\theta^{\rho\sigma}{\cal
P}_\rho\otimes{\cal P}_\sigma}[{\hat\phi}(x')\otimes{\hat\phi}(y')
\]
\[ -\ \frac{\rm i}{2}\theta^{\rho\sigma}{\omega_\rho}^\nu({\cal
P}_\nu{\hat\phi})(x')\otimes({\cal P}_\sigma{\hat\phi})(y')\ -\
\frac{\rm i}{2}\theta^{\rho\sigma} {\omega_\sigma}^\nu ({\cal
P}_\rho {\hat\phi})(x')\otimes ({\cal P}_\nu{\hat\phi}) (y')]\}\]
\[ \doteq\ m\{e^{-\frac{\rm i}{2}\theta^{\rho\sigma}({\cal
P}_\rho+{\omega_\rho}^\nu{\cal P}_\nu)\otimes({\cal
P}_\sigma+{\omega_\sigma}^\nu{\cal
P}_\nu)}({\hat\phi}(x')\otimes{\hat\phi}(y'))\}\]
\[ =\ e^{\frac{\rm i}{2}\theta^{\rho\sigma}
\partial_{x^{\prime\rho}}\partial_{y{\prime^\sigma}}}{\hat\phi}(x')
{\hat\phi}(y')\ =\ {\hat\phi}(x')\star{\hat\phi}(y')\ .\]
The symbol $\doteq$ denotes an equality up to first order in
$\omega$; $x^{\prime\rho}=x^\rho-{\omega_\mu}^\rho x^\mu$ and
$y^{\prime\sigma}=y^\sigma-{\omega_\mu}^\sigma y^\mu$ are just the
Lorentz transformed space-time points.

This proves the covariance relation (\ref{I.8}) for infinitesimal
Lorentz transformation. For a finite transformation, (\ref{I.8})
is recovered by a usual exponentiation. The inclusion of
translations is trivial. We point out that the covariance relation
can be
extended to multiple products of field.

\end{document}